# Elastic Constant Measurement from Vibrational Mode Frequencies in Resonant Ultrasound Spectroscopy

Barnana Pal

Saha Institute of nuclear Physics,
Condensed Matter Physics Division,
1/AF, Bidhannagar, Kolkata-700064, India.
e-mail: barnana.pal@saha.ac.in

Abstract:

Execution of Resonant Ultrasound Spectroscopy (RUS) for accurate measurement of elastic constants lies primarily on a perfect matching in the calculated and measured mode frequencies of free vibration. Calculation of these frequencies require estimated values of the elastic constants of the material under study, and one has to depend on other experiments for these data. The present work proposes and demonstrates an alternative to derive initial guess values of essential parameters for an isotropic and homogeneous material from the acquired RUS spectra itself. Specimen samples are taken in the shape of rectangular parallelepiped having nearly same cross-sectional dimension but with different lengths. For particular compressional (shear) mode corresponding to length $l$, the frequency $f$ is inversely proportional to $l$. The slope $m$ of $f$ versus $1/l$ plot equals to half of the compressional (shear) velocity and this in turn gives an estimate of $c_{11}$ ($c_{44}$). With these parameters as the input guess parameters, RUS fitting method is executed to find out the best fit results. Elastic constants of commercially available specimens of aluminium, copper, lead, steel and brass are measured in this technique assuming macroscopic homogeneity. Results show good agreement with available literature values.

## 1. Introduction:

Determination of elastic constants in specimens of relatively small size, dimension ~ a few mm³, has been made possible with Resonant Ultrasound Spectroscopy (RUS) [1-6]. The normal modes of free vibrations of the sample are excited using a transducer oscillating at a fixed frequency and another sweep frequency transducer is used to identify the resonance mode frequencies. The normal mode frequencies are characteristics of the size, shape, density and elastic constants of the material. These frequencies are calculated using initial guess values of the elastic constants. The success of RUS technique depends on the matching of the measured frequencies and the calculated mode frequencies. The matching is done by an iterative method by adjusting the values of the elastic constants. An improper guess for the elastic parameters either lead to erroneous results or failure of the iterative process towards convergence. The prime disadvantage of the RUS method lies in the fact that one has to depend on other experiments for initial guess values of elastic constants. In this report a method is proposed to extract these guess parameters from the acquired RUS spectra itself.

Commercially available aluminium, copper, lead, brass and steel are taken as test samples. The samples are polycrystalline. In general, polycrystalline samples are aggregates of crystal grains of different sizes and shapes situated with random orientations inside the material. The macroscopic behaviour of these materials may be considered as isotropic and homogeneous and their elastic properties can be obtained from single crystal properties by averaging over all possible grain sizes and crystal orientations. Voigt [7] and Reuss [8] obtained average values for polycrystalline substances from respective single crystal constants. Hill [9] showed that these averages represent the lowest upper bound and highest lower bound respectively and are termed as Voigt and Reuss averages. Subsequently Hashin and Shtrikman [10, 11] improved the expressions for Voigt and Reuss bounds for cubic crystals. These are known as Haskin (upper) bound and Shtrikman (lower) bound. The results of present experiment are compared with these average values [12] and other reported values [13, 14].

The background theory is outlined in sec. 2, experimental method and results along with available literature values are presented in sce. 3, results are discussed in sec. 4 and sec. 5 gives concluding remarks.

## 2. Background Theory:

For an isotropic solid, the elastic property is completely described by Lamé constants λ and μ. The elastic tensor $c_{\alpha\beta}$ is given by,

$$c_{\alpha\beta} = \begin{pmatrix} \lambda + 2\mu & \lambda & \lambda & 0 & 0 & 0 \\ \lambda & \lambda + 2\mu & \lambda & 0 & 0 & 0 \\ \lambda & \lambda & \lambda + 2\mu & 0 & 0 & 0 \\ 0 & 0 & 0 & \mu & 0 & 0 \\ 0 & 0 & 0 & 0 & \mu & 0 \\ 0 & 0 & 0 & 0 & 0 & \mu \end{pmatrix}$$

$c_{11} = \lambda + 2\mu$ is compressional modulus and $c_{44} = \mu$ is shear modulus. Resonant mode frequencies for free vibration depends on sample shape, size, density (ρ), and the constants $c_{\alpha\beta}$. In the present experiment, samples are taken in the shape of a rectangular parallelepiped of length $l$, breadth $b$ and height $h$. Among the various possible vibrational modes, there are compressional modes whose frequencies ($f$) are integral multiples of $\sqrt{c_{11}/\rho}/2d$ and shear modes having frequencies integral multiples of $\sqrt{c_{44}/\rho}/2d$. Here $d$ may be any one of $l$, $b$ and $h$. Thus, for a particular shear or compressional mode, $f$ vs. $1/d$ plot will be a straight line. The gradient is $m_1 = \sqrt{c_{44}/\rho}/2$ for shear mode frequencies and for compressional mode frequencies it is $m_2 = \sqrt{c_{11}/\rho}/2$. Thus, identification of proper mode frequencies enables one to find out $c_{11}$ and $c_{44}$ from the RUS spectra acquired for different sample lengths.

## 3. Experimental Method and Results:

The experiment is done using Magnaflux Quasar Resonant Ultrasound Spectroscopy system. The samples are taken in the shape of rectangular parallelepiped having nearly same cross-sectional dimension but with varying lengths. The resonance frequency modes are detected starting from the lowest frequency, for at least three samples having different lengths ($l$). The frequency $f$ is plotted against $1/l$. For a particular sample, the frequency points will lie on a vertical line. For larger $l$, lower will be the mode frequencies. Further, the lowest frequency mode is expected to be a shear mode. In the $f$ vs. $1/l$ plot the lowest frequency points for different samples will lie on a straight line and $c_{44} = \mu$ can be determined from the slope $m_1$ of this line. For compressional modes difficulty arises in identifying the mode frequencies due to overlap with other possible mode frequencies. However, a few guess values may be found out by identifying frequency modes lying on a straight line with slope $m_2 > \sqrt{2} \, m_1$ and any of these may be a logical guess for $m_2$ or $c_{11}$. Each of these sets of $c_{44}$ and $c_{11}$ may be considered as an input to run the RPR fitting program in the RUS to get the final output parameters with minimum $\chi^2$ and rms error.

The dimensions and densities $\rho$ of the samples under study are listed in Table-I. The purpose of the present work is to demonstrate the applicability of RUS for any sample even if there is no idea about its elastic constants. The acquired RUS spectra are good enough in estimating input guess parameters to run the RPR fitting code. As such, commercially available samples are chosen without going to the minute compositional details and crystalline properties.

*Table-I: Test sample densities in gm/cc and dimensions in cm.*

| Sample | Average density $\rho$ (gm/cc) | Specimen no. | Length ($l$) (cm) | Breadth ($b$) (cm) | Height ($h$) (cm) |
|---|---|---|---|---|---|
| Al | 2.56 | Al1 | 0.700 | 0.282 | 0.220 |
|  |  | Al2 | 0.598 | 0.296 | 0.236 |
|  |  | Al3 | 0.500 | 0.284 | 0.216 |
| Cu | 8.71 | Cu1 | 0.698 | 0.300 | 0.232 |
|  |  | Cu2 | 0.600 | 0.296 | 0.232 |
|  |  | Cu3 | 0.500 | 0.294 | 0.232 |
|  |  | Cu4 | 0.398 | 0.298 | 0.232 |
|  |  | Cu5 | 0.350 | 0.298 | 0.230 |
| Pb | 10.82 | Pb1 | 0.658 | 0.348 | 0.314 |
|  |  | Pb2 | 0.600 | 0.350 | 0.314 |
|  |  | Pb3 | 0.550 | 0.340 | 0.318 |
|  |  | Pb4 | 0.506 | 0.350 | 0.314 |
|  |  | Pb5 | 0.450 | 0.348 | 0.312 |
| Steel | 7.50 | St1 | 0.698 | 0.296 | 0.228 |
|  |  | St2 | 0.598 | 0.290 | 0.226 |
|  |  | St3 | 0.494 | 0.294 | 0.226 |
|  |  | St4 | 0.396 | 0.264 | 0.220 |
|  |  | St5 | 0.344 | 0.284 | 0.226 |
| Brass | 8.21 | Br1 | 0.648 | 0.340 | 0.270 |
|  |  | Br2 | 0.598 | 0.340 | 0.270 |
|  |  | Br3 | 0.546 | 0.342 | 0.270 |
|  |  | Br4 | 0.494 | 0.340 | 0.270 |
|  |  | Br5 | 0.450 | 0.340 | 0.270 |

The $f$ vs. $1/l$ plot for the samples Al, Cu, Pb, brass and steel are presented in figures 1,2,3,4 and 5 respectively. The mode frequencies detected for a particular sample with length $l$ lies on a vertical line at x = $1/l$. In each figure three lines are drawn. These are the best linear fits for the points shown in larger point size. The line LF1, with lowest slope $m_1$ is the linear fit for lowest frequency points. These should correspond to shear modes and shear modulus $c_{44} = \mu$ is determined from $m_1$. LF2 and LF3 represent linear fits for possible compressional mode points giving rise to two possible guess values for $m_2$ and hence for compressional modulus $c_{11} = \lambda + 2\mu$. The guess values thus obtained are tabulated in Table-II. Each set of these guess values $c_{44}$ and $c_{11}$ is used as input parameters to run the fit parameter program for all of the acquired frequency spectra for a specific sample and the output with minimum $\chi^2$ and rms error is accepted as the final result. The results are presented in Table-III along with other experimental values available in the literature.

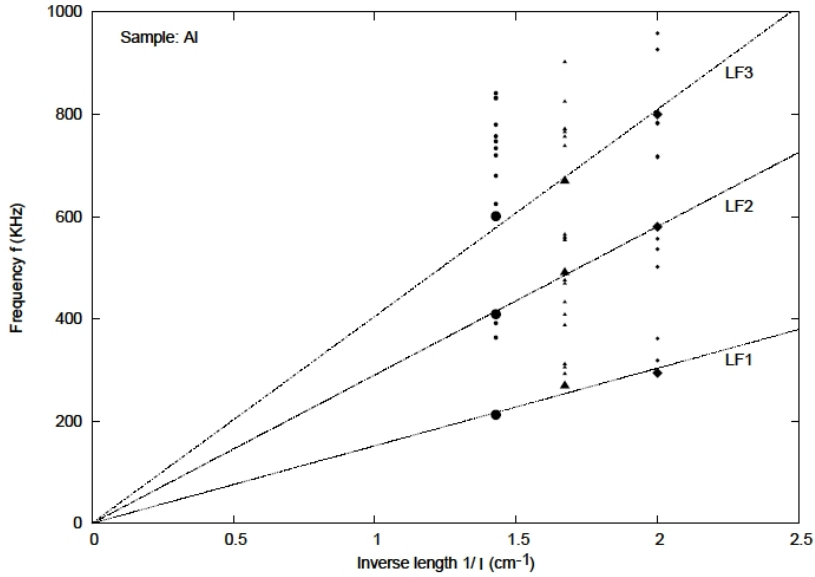

*Figure 1. f (in KHz) vs. 1/l (in cm$^{-1}$) plot for Al. LF1 is the linear fit for lowest frequency points representing possible shear modes. Two possible linear fits for compressional mode frequencies are LF2 and LF3. For each of the three lines, the frequency points are shown in larger point size.*

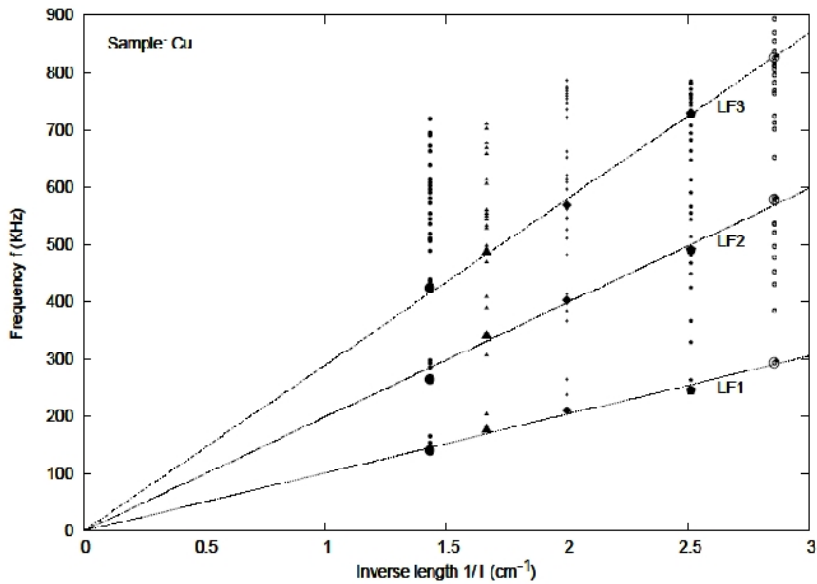

*Figure 2. f (in KHz) vs. 1/l (in cm$^{-1}$) plot for Cu. LF1 is the linear fit for lowest frequency points representing possible shear modes. Two possible linear fits for compressional mode frequencies are LF2 and LF3. For each of the three lines, the frequency points are shown in larger point size.*

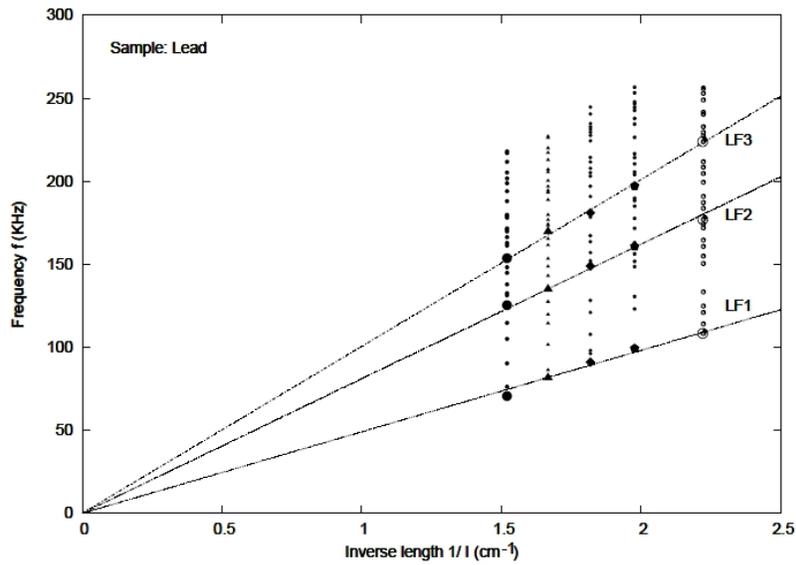

*Figure 3. f (in KHz) vs. 1/l (in cm$^{-1}$) plot for Pb. LF1 is the linear fit for lowest frequency points representing possible shear modes. Two possible linear fits for compressional mode frequencies are LF2 and LF3. For each of the three lines, the frequency points are shown in larger point size.*

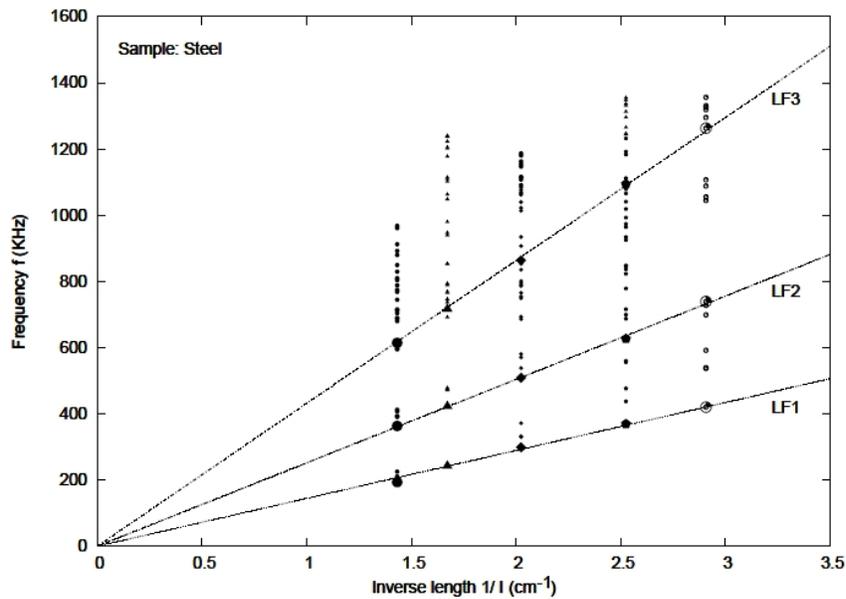

*Figure 4. f (in KHz) vs. 1/l (in cm$^{-1}$) plot for steel. LF1 is the linear fit for lowest frequency points representing possible shear modes. Two possible linear fits for compressional mode frequencies are LF2 and LF3. For each of the three lines, the frequency points are shown in larger point size.*

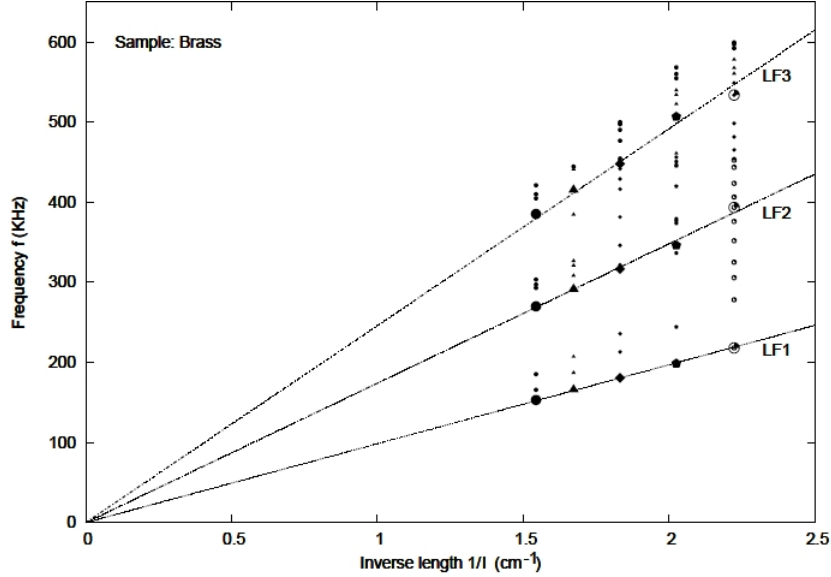

*Figure 5. $f$ (in KHz) vs. $1/l$ (in cm$^{-1}$) plot for brass. LF1 is the linear fit for lowest frequency points representing possible shear modes. Two possible linear fits for compressional mode frequencies are LF2 and LF3. For each of the three lines, the frequency points are shown in larger point size.*

*Table-II: Guess values of $c_{44}$ and $c_{11}$ in GPa calculated from the slopes of linear fits LF1, LF2 (option1) and LF3 (option2) in $f$ vs. $1/l$ plot presented in figures 1,2,3,4 and 5 for Al, Cu, Pb, steel and brass respectively.*

| Sample | Guess Parameters | | | | | | |
|---|---|---|---|---|---|---|---|
| | Density $\rho$ in gm/cc | $m_1$ in Km/s | $c_{44}$ Gpa | Option 1 | | Option 2 | |
| | | | | $m_2$ in Km/s | $c_{11}$ Gpa | $m_2$ in Km/s | $c_{11}$ Gpa |
| Al | 2.56 | 1.52 | 23.66 | 2.90 | 86.12 | 4.05 | 167.96 |
| Cu | 8.71 | 1.02 | 36.25 | 1.99 | 137.97 | 5.80 | 293.00 |
| Pb | 10.82 | 0.49 | 10.43 | 0.81 | 28.46 | 1.01 | 43.76 |
| Steel | 7.50 | 1.45 | 63.08 | 2.52 | 190.51 | 4.32 | 559.87 |
| Brass | 8.21 | 0.99 | 32.19 | 1.74 | 99.43 | 2.46 | 198.73 |

## 4. Discussion:

The slopes of linear fits LF1, LF2 and LF3 (fig. 1) for Al sample are listed in Table-II. These are 1.52 Km/s, 2.90 Km/s and 4.05 Km/s respectively. Slope ($m_1$) of LF1 gives guess value for $c_{44}$ which is 23.66 GPa and slopes ($m_2$) of LF2 and LF3 give two optional guess values for $c_{11}$, 86.12 GPa and 167.96 GPa. With these guess parameters as input, RPR fitting code is run for each of the three spectra acquired for three Al samples having nearly same cross-sections but different lengths. The best fit output parameters with minimum rms error (less than 1.0) is accepted as the final result which are tabulated in Table-III. These are: $c_{11}$ = 103.85 GPa, $c_{44}$ = 23.32 GPa, E = 63.20 GPa, $\nu$ = 0.355, $u_c$ = 6.326 Km/s, $u_s$=2.998 Km/s. The experimental results are compared with calculated elastic properties for random, macroscopically isotropic aggregates of crystalline substances available in the literature [12]. It is to be noted that, except v, all of the parameter values are a little less than the lower limit of the predicted range and the deviations are ≤ 10%. The value of v is higher than the predicted upper limit and the deviation is 2%.

*Table-III: RUS best-fit results for compressional modulus $c_{11}$ in GPa, shear modulus $c_{44}$ in GPa, Young's modulus E in GPa, Poisson's ratio v, compressional wave velocity $u_c$ in Km/s, and shear wave velocity $u_s$ in Km/s compared with literature data. Percentage deviation is calculated using lower(upper) limiting value in case the experimental value is smaller(greater) than lower(upper) limit.*

| Sample | Measured parameter | Present expt. | Literature data | Reference | Deviation % |
|---|---|---|---|---|---|
| Al | $c_{11}$ (GPa) | 103.85 | - | | |
| | $c_{44}$ (GPa) | 23.32 | 26.0-26.3 | 12 | 10 |
| | E (GPa) | 63.20 | 70.1-70.7 | 12 | 10 |
| | v | 0.355 | 0.346-0.347 | 12 | 2 |
| | $u_c$ (Km/s) | 6.326 | 6.414-6.424 | 12 | 1 |
| | $u_s$ (Km/s) | 2.998 | 3.104-3.119 | 12 | 3 |
| Cu | $c_{11}$ (GPa) | 173.21 | - | | |
| | $c_{44}$ (GPa) | 47.19 | 40.1-54.6 | 12 | Within range |
| | E (GPa) | 123.89 | 109.8-144.8 | 12 | Within range |
| | v | 0.313 | 0.326-0.368 | 12 | 4 |
| | $u_c$ (Km/s) | 4.449 | 4.638-4.865 | 12 | 4 |
| | $u_s$ (Km/s) | 2.322 | 2.121-2.475 | 12 | Within range |
| Pb | $c_{11}$ (GPa) | 25.91 | - | | |
| | $c_{44}$ (GPa) | 11.04 | 6.7-10.1 | 12 | 9 |
| | E (GPa) | 24.93 | 19.0-28.1 | 12 | Within range |
| | v | 0.129 | 0.388-0.424 | 12 | 67 |
| | $u_c$ (Km/s) | 1.552 | 2.111-2.205 | 12 | 26 |
| | $u_s$ (Km/s) | 1.013 | 0.767-0.945 | 12 | 7 |
| Steel | $c_{11}$ (GPa) | 246.71 | - | | |
| | $c_{44}$ (GPa) | 79.99 | 72-89 | 13 | Within range |
| | E (GPa) | 201.60 | 188-224 | 13 | Within range |
| | v | 0.260 | 0.270-0.329 | 13 | 4 |
| | $u_c$ (Km/s) | 5.691 | 5.759 | 14 | 1 |
| | $u_s$ (Km/s) | 3.240 | 3.134 | 14 | 3 |
| Brass | $c_{11}$ (GPa) | 164.05 | - | | |
| | $c_{44}$ (GPa) | 39.25 | 21.4-107.5 | 12 | Within range |
| | E (GPa) | 105.41 | 54.5-169.5 | 12 | Within range |
| | v | 0.343 | -0.212-0.271 | 12 | 27 |
| | $u_c$ (Km/s) | 4.478 | 2.859-4.682 | 12 | Within range |
| | $u_s$ (Km/s) | 2.190 | 1.602-3.588 | 12 | Within range |

The slopes of LF1, LF2 and LF3 for Cu sample (fig 2) as given in Table-II are 1.02 Km/s, 1.99 Km/s and 5.80 Km/s respectively. Guess value of $c_{44}$ is obtained as 36.25 GPa and two guess values of $c_{11}$ are calculated as 137.97 GPa and 293.00 GPa. The best fit output parameters as presented in Table-III are: $c_{11}$ = 173.21 GPa, $c_{44}$ = 47.19 GPa, E = 123.89 GPa, v = 0.313, $u_c$ = 4.449 Km/s, $u_s$ = 2.322 Km/s. Comparing with literature data [12] it is seen that all of the constants are within the predicted range except ν and u_c. These two parameters are a little lower than the lower limit of the predicted range. For each of these two the deviation is 4%.

From fig. 3, the slopes LF1, LF2 and LF3 for Pb are obtained as 0.49 Km/s, 0.81 Km/s and 1.01 Km/s respectively. Table-II shows these values. Corresponding guess parameter $c_{44}$ is 10.43 GPa and two guess values of $c_{11}$ are 28.46 GPa and 43.76 Gpa. Best fit output parameters given in Table-III are: $c_{11}$ = 25.91 GPa, $c_{44}$ = 11.04 GPa, E = 24.93 GPa, ν = 0.129, $u_c$ = 1.552 Km/s, $u_s$ = 1.013 Km/s. Comparison with literature data [12] shows that, for this sample $c_{44}$ is a little higher than the upper limit of predicted range, deviation is 9%, E is within the predicted range, ν and $u_c$ are reasonably less than the lower limit of prescribed limiting values and $u_s$ is a little higher than the upper limiting value. For ν, the deviation is 67%, for $u_c$ the deviation is 26% and for $u_s$ the deviation is 7%

Fig. 4 gives the slopes LF1, LF2 and LF3 for steel. Table-II shows that these are 1.45 Km/s, 2.52 Km/s and 4.32 Km/s respectively. Corresponding guess value $c_{44}$ is 63.08 GPa and two guess values of $c_{11}$ are 190.51 GPa and 559.87 Gpa. Best fit output parameters obtained in this case as seen in Table-III are: $c_{11}$ = 246.71 GPa, $c_{44}$ = 79.99 GPa, E = 201.60 GPa, ν = 0.260, $u_c$ = 5.691 Km/s, $u_s$ = 3.240 Km/s. These are in very good agreement with literature data [13,14]. The parameters $c_{44}$ and E are within predicted limiting values. ν is a little lower than the predicted lower limit, the deviation being less than 4%. $u_c$ and $u_s$ agree well with the literature value.

For brass the slopes LF1, LF2 and LF3 as obtained from fig. 5 are 0.99 Km/s, 1.74 Km/s and 2.46 Km/s respectively. Calculated guess parameter $c_{44}$ is 32.19 GPa and two guess values of $c_{11}$ are 99.43 GPa and 198.73 Gpa. Best fit results for brass are: $c_{11}$ = 164.05 GPa, $c_{44}$ = 39.25 GPa, E = 105.41 GPa, ν = 0.343, $u_c$ = 4.478 Km/s, $u_s$ = 2.190 Km/s. Comparison with literature data [12] given in Table-III shows that all of the parameter values are well within the predicted range except ν, which is higher than the upper limiting value, the deviation being 27%.

Among the five samples studied, agreement between experimental and literature data are very good for steel and Cu, and reasonably good for brass. For Al and Pb the agreement is not so satisfactory. The reason may be their soft and malleable nature. Sample hardness is one of the important criteria for RUS to get good result. The calculated mode frequencies depend on sample shape and dimension. For soft samples there is every possibility of shape deformation in handling them during experiment. As a consequence, calculated mode frequencies will differ considerably from their actual values leaving scope for convergence towards inappropriate final parameter values.

Another important point to note is that, the output value of shear modulus $c_{44}$ does not differ too much from its guess value determined from the $f$ vs. $1/l$ plot. The difference is within 25% of the final result. Since the shear mode is low frequency mode, it is relatively easy to identify these for varying sample lengths and then finding out the linear fit LF1. On the other hand, longitudinal modes are mixed up with other possible modes creating uncertainty in identifying the right frequencies and consequently finding linear fits LF2 or LF3. However this is not a serious issue, at least for the present study. The guess parameters determined from the $f$ vs. $1/l$ plot are found good enough to run the RPR fitting program towards convergence.

## 4. Conclusion:

In the present study a method for the extraction of input guess values for the elastic constants from the acquired RUS spectra itself has been proposed. It is shown that guess parameters $c_{44}$ and $c_{11}$ can be calculated from $f$ vs. $1/l$ plot. Not only that, the final results show that $c_{44}$ obtained in this method is a reasonably good guess. However, confusion arises in estimating $c_{11}$, but this does not appear to be a serious issue. Among the five samples studied in this method, very good results are obtained for Cu and steel, reasonably good results are obtained for brass and results for Al and Pb are not so satisfactory. The reason is attributed to the softness and malleability of Al and Pb samples that may cause shape deformation in handling the samples during experiment.

**Acknowledgement:**

The author is indebted to Dr. Jishnu Basu, Mr Sudipta Barman, and Mr Supriyo Barman for sample preparation and to Mrs Papia Mondal and Mrs Sankari Chakraborty for necessary technical help.